\documentstyle[12pt]{article}
\begin{document}
\begin{titlepage}
\begin{flushright}
UAB--FT--398\\
L.P.T.H.E.--ORSAY 96/73\\
September 1996\\
hep-ph/9609423
\end{flushright}
\vspace{1cm}

\begin{center}
{\bf\LARGE Improved Bounds on the Electromagnetic\\[0.6ex]
 Dipole Moments of the Tau Lepton}
\vspace{2cm}

\centerline{\bf R. Escribano$^{a}$ and
		E. Mass\'o$^{a,b}$
}
\vspace{0.6cm}

\centerline{$^a$Grup de F\'{\i}sica Te\`orica and IFAE,
		Universitat Aut\`onoma de Barcelona,}
\centerline{E-08193 Bellaterra, Spain}
\centerline{$^b$LPTHE, b\^at. 211, Universit\'e Paris XI, Orsay Cedex, 
France$^{\ast}$}
\vspace{3cm}

{\bf Abstract}\\
\parbox[t]{\textwidth}
{Using electroweak data and an effective Lagrangian approach we obtain 
stringent bounds on the tau anomalous magnetic moment, 
$-0.004 \leq a_\tau \leq 0.006$, and on its electric dipole moment, 
$|d_\tau| \leq 1.1 \times 10^{-17}\ e$\,cm.
This significantly improves our previous bounds.
With the same method, we obtain limits on the dipole moments of other 
fermions.}
\end{center}
\vspace{\fill}

{\noindent\makebox[10cm]{\hrulefill}\\
\footnotesize
\makebox[1cm][r]{$^{\ast}$}Laboratoire associ\'e au C.N.R.S. --- URA D0063.
}
\end{titlepage}
\newpage
 
\section{Introduction}

Tau lepton physics (see Ref.'s \cite{MPerl} and the report of 
K. Hayes in \cite{KHayes}) provides a laboratory to test ideas and models 
beyond the Standard Model such as compositeness and/or new sources of 
CP-violation. 

In particular, the tau magnetic and electric dipole moments may play an
important role in elucidating the nature of this heavy lepton. From the 
gene\-ral electromagnetic matrix element describing the interaction of a
$\tau$ lepton with photons
\begin{equation}
\label{matrixelem}
\langle p' | J_{em}^\mu (0) | p \rangle = 
e \, \bar u(p') \left(  F_1 \, \gamma^\mu + 
\left[  \frac{i}{2 m_\tau} \, F_2 +  \gamma_5 \, \tilde F_2 \right] 
\sigma^{\mu \nu} \, q_\nu   
\right) u(p) \ ,
\end{equation}
where $q=p'-p$ and $F_1(q^2=0)=1$, 
one defines the tau anomalous magnetic dipole moment
\begin{equation}
\label{atauteo}
a_\tau = \frac{g_\tau-2}{2} = F_2(q^2=0)\ ,
\end{equation}
and the tau electric dipole moment (EDM)
\begin{equation}
\label{dtauteo}
d_\tau = e \, \tilde F_2(q^2=0) \ .
\end{equation}

In the Standard Model (SM), the tau is considered to be a spin-1/2
pointlike Dirac particle. A theoretical SM calculation \cite{Samuel} predicts 
the value
\begin{equation}
\label{atauSM}
a_\tau |_{SM} = 1.177 \times 10^{-3}\ ,
\end{equation}
and a very small value for $d_\tau$,
the precise value depending on unknown leptonic mixing angles
and neutrino masses \cite{Donoghue}.

Up to now, the most stringent way to constrain the electromagnetic
dipole moments $a_\tau$ and $d_\tau$ has shown to be the use of the
method we developed in Ref.\,\cite{rem}. 
The idea is the following. The addition
of new physics to the SM could lead to values for $a_\tau$ different
from the value of Eq.\,(\ref{atauSM}) or to $d_\tau$ being larger
than the SM predicts. We describe
potential deviations from the SM induced by some kind of 
new physics by a set of effective Lagrangians that
contain gauge-invariant operators contributing to the tau dipole moments. 
Such effective Lagrangians inevitably induce anomalous couplings of the
$Z$ to the tau lepton, which would contribute to the partial width
$\Gamma_\tau \equiv \Gamma ( Z \rightarrow \tau \bar\tau )$.
The tight constraints of LEP1 data on deviations from the SM can 
be used to get model independent bounds on the electromagnetic dipole 
moments.  

It is well known that the impressive accuracy of the LEP1
measurements makes necessary the inclusion of radiative corrections when
comparing theory and experiment.
The theoretical SM prediction for $\Gamma_\tau$ ---which is an input in
our analysis--- depends on the top quark and Higgs masses, $m_t$ and
$m_H$. When we worked out the above mentioned idea in 1992 there was 
only an experimental lower bound on $m_t$ \cite{CDF}; therefore a
two-parameter analysis was inevitable. (One of the parameters was $m_t$
and the other took into account the presence of new physics.) 
In the meanwhile, evidence for the top quark \cite{CDF&D0ultim}
has shown up and the corresponding $m_t$ measurements \cite{mtCDF&D0}
allow us to conside\-rably improve our previous bounds by directly comparing
the SM prediction for $\Gamma_\tau$ with its experimental value. In
addition, we now have more data at our disposal including the 1995 
run \cite{LEPEWWG}. The substantial
increase in the  precision of the experimental measurements also 
improves our bounds. The main purpose of this
letter is to show that the bounds on $a_\tau$ and $d_\tau$ are tightened
significantly  when we use the same ideas as \cite{rem}
but employ the new experimental data. Using the same method,
we find upper limits on the dipole moments of some other fermions. 
In this case we also achieve an improvement with respect to the limits
presented in Ref.\,\cite{rem2}.

In this letter, we will show the formulae
obtained in the framework of a linearly realized effective Lagrangian. 
In Ref.\,\cite{rem2} it was shown that a non-linearly realized effective
Lagrangian would lead to the same numerical limits for the dipole moments.
    
\section{Bounds on tau lepton dipole moments}

The lowest dimension gauge-invariant operators built from the SM fields 
that contribute to the tau anomalous magnetic dipole moment $a_\tau$ are:
\begin{equation}
\label{opefftauB}
{\cal O}_{\tau B} = 
\overline{L_\tau} \, \sigma^{\mu \nu} \, \tau_R \,
\Phi \, B_{\mu \nu} \ ,
\end{equation}
and
\begin{equation}
\label{opefftauW}
{\cal O}_{\tau W} = 
\overline{L_\tau} \, \sigma^{\mu \nu} \, \vec {\bf \sigma} \, \tau_R \,
\Phi \, \vec  W_{\mu \nu} \ .
\end{equation}
Here, $L_\tau$ is the $\tau$ left-handed
isodoublet, $\tau_R$ its right-handed partner, $\vec W, B$ are the SU(2) and
U(1) field strengths, and $\Phi$ is the scalar doublet. Contributions to
$a_\tau$ appear below the weak scale due to spontaneous symmetry breaking,
\begin{equation}
\label{Phi}
\Phi \rightarrow \left( 
\begin{array}{c}
0 \\
v/\sqrt 2
\end{array}
\right) + \ldots
\end{equation}
with $v^2 = 1/(\sqrt 2 \, G_F) \simeq (246 \, GeV)^2$.

Let us examine the consequences of the operator 
$O_{\tau B}$ in (\ref{opefftauB}).
An effective Lagrangian containing this operator, namely
\begin{equation}
\label{lefftauB}
\delta {\cal L}= {\alpha_{\tau B} \over \Lambda^2}\,
{\cal O}_{\tau B}\ ,
\end{equation}
where an unknown coupling constant $\alpha_{\tau B}$ and
an unknown high-energy mass scale $\Lambda$ are introduced, would contribute
to $a_\tau$ as
\begin{equation}
\label{atau}
a_\tau=a_\tau|_{SM}+\delta a_\tau\ ,
\end{equation}
and simultaneously would contribute to $\Gamma_\tau$ as
\begin{equation}
\label{gammatau}
\Gamma_\tau=\Gamma_\tau|_{SM}+\delta \Gamma_\tau\ .
\end{equation}
Both contributions are related in the following manner:
\begin{equation}
\label{ataugammatau}
{ e \over 2  m_\tau }\, |\delta a_\tau|= \cot \theta_w\, 
\left[{24\, \pi \over M_Z^3\, \sqrt{1-4\left({m_\tau \over M_Z}\right)^2}}\, 
|\delta \Gamma_\tau| \right]^{1/2}\ ,
\end{equation}
where $\theta_w$ is the weak mixing angle. 
It is important to realize that the relation 
(\ref{ataugammatau}) is independent of $\alpha_{\tau B}$ and
$\Lambda$. This fact allows us to place a bound on $a_\tau$
in a model independent way.

The SM theoretical prediction is given by \cite{bhm}
\begin{equation}
\label{GammatauSM}
\Gamma_\tau|_{SM}=83.78\pm 0.05\pm 0.05\ \mbox{MeV}\ ,
\end{equation}
where the first error comes from using the interval $m_t = 175\pm 6$ GeV 
for the top mass \cite{mtCDF&D0} and the second corresponds to the 
interval $m_H = 150^{+150}_{-90}$ GeV for the Higgs mass \cite{LEPEWWG}.
To be conservative, we add these two ``theoretical errors'' linearly.
The prediction (\ref{GammatauSM}) has to be compared with the experimental
value \cite{LEPEWWG}
\begin{equation}
\label{Gammatauexp}
\Gamma_\tau|_{exp}=83.72\pm 0.26\ \mbox{MeV}\ .
\end{equation}
The agreement between experiment and the SM prediction for
$\Gamma_\tau$ restricts potential new physics contributions to 
the tau anomalous magnetic moment. At the 95$\%$ CL we find the upper bound
\begin{equation}
\label{ataueff}
|\delta a_\tau | \leq 0.0048 \ . 
\end{equation}

The same analysis can be done using now the operator in (\ref{opefftauW}).
We get a stronger upper bound on $\delta a_\tau$ (by a factor of about 3). 
Thus, we conclude that our bound on the tau
anomalous magnetic moment is the one given in (\ref{ataueff}). 
In so doing, we exclude unnatural cancellations
between the effects of the two independent operators ${\cal O}_{\tau B}$ and 
${\cal O}_{\tau W}$.

Combining (\ref{ataueff}) with the SM prediction (\ref{atauSM}),
we finally obtain our 95$\%$~CL bound on the tau magnetic dipole moment
\begin{equation}
\label{ataufinal}
-0.004\leq  a_\tau \leq 0.006 \ \ \ \ (2\sigma ) \ ,
\end{equation}
which improves our previous limit \cite{rem} by a factor of about two.

Now we turn our attention to the tau EDM, $d_\tau$.
There are also two relevant operators
\begin{equation}
\label{opeffttauBtilde}
\tilde{\cal O}_{\tau B} =  
\overline{L_\tau} \, \sigma^{\mu \nu} \, i\gamma_5 \, \tau_R \,
\Phi \, B_{\mu \nu} \ ,
\end{equation}
and
\begin{equation}
\label{opeffttauWtilde}
\tilde{\cal O}_{\tau W} =
\overline{L_\tau} \, \sigma^{\mu \nu} \, i\gamma_5 \, \vec {\bf \sigma} \, 
\tau_R \, \Phi \, \vec W_{\mu \nu} \ . 
\end{equation}
To place a bound on $d_\tau$, we could follow a similar procedure
as we followed for $a_\tau$. Indeed, an effective Lagrangian 
containing the operator $\tilde{\cal O}_{\tau B}$
\begin{equation}
\label{lefftauBtilde}
\delta {\cal L}= {\tilde \alpha_{\tau B} \over {\tilde \Lambda}^2}\,
\tilde{\cal O}_{\tau B}\ ,
\end{equation}
contributes to the EDM and modifies $\Gamma_\tau$. The limits on
$\Gamma_\tau$ imply the bound 
$|d_\tau| \leq 2.7 \times 10^{-17}\ e$\,cm (95$\%$ CL).

However, we get a much better bound if we use the results 
from the search for
CP violation in the decay $Z \rightarrow \tau \bar \tau$ at
LEP \cite{limitsCP,millorlimitCP}. Effects
coming from a possible CP-violating $Z \tau \bar \tau$ vertex
\begin{equation}
\label{matrixelemZ}
\langle p' | J_Z^\mu (0) | p \rangle =
e\, \bar u(p')\, \tilde F^w_2\,
\sigma^{\mu \nu}\, \gamma_5\, q_\nu\, u(p)\ , 
\end{equation}
have not been observed. This negative result implies a limit on the
weak dipole moment of the tau lepton, $d^w_\tau$, defined as
\begin{equation}
\label{dw}
d^w_\tau = e\, \tilde F^w_2 (q^2 = M_Z^2) \ .
\end{equation}

Using the effective Lagrangian (\ref{lefftauBtilde}) yields
a simple relationship between the electric and the weak dipole moments:
\begin{equation}
\label{dedw}
|d_\tau| = \cot \theta_w \, |d^w_\tau| \ .
\end{equation} 

The most stringent limit obtained at LEP for $d^w_\tau$ 
is \cite{millorlimitCP}  
$|d^w_\tau| \leq 5.8 \times 10^{-18}\ e$\,cm at 95\% CL. 
Using now (\ref{dedw}), we get the following upper limit on the tau EDM
\begin{equation}
\label{dtaufinal}
|d_\tau| \leq 1.1 \times 10^{-17}\ e\,\mbox{cm} \ \ \ \ (2\sigma) \ ,  
\end{equation}
which improves our previous bound \cite{rem} by a factor of about four.

The effective Lagrangian built with the second operator in
(\ref{opeffttauWtilde}) leads to a stronger limit on $d_\tau$ so that
we keep the bound in (\ref{dtaufinal}) as our limit. In so doing, we do not
allow for unnatural cancellations between the operators
(\ref{opeffttauBtilde}) and (\ref{opeffttauWtilde}).

\section{Bounds on other fermions dipole moments}

The procedure used to constrain $a_\tau$ can obviously be
extended to other fermions. For the magnetic moment of the tau 
neutrino we get an upper bound 
\begin{equation}
\label{anutau}
|\mu(\nu_\tau)| \leq 2.7 \times 10^{-6}\ \mu_B\ \ \ \  (2\sigma) \ ,
\end{equation}
where $\mu_B = e/2 m_e$ is the Bohr magneton. For the quarks, we obtain
(at 95\%~CL)
\begin{equation}
\label{aquarks}
\begin{array}{l}
|a_u| \leq 1.8 \times 10^{-5}\ ,\\[1ex]
|a_d| \leq 7.3 \times 10^{-5}\ ,\\[1ex]
|a_s| \leq 1.5 \times 10^{-3}\ ,\\[1ex]
|a_c| \leq 4.8 \times 10^{-3}\ ,\\[1ex]
|a_b| \leq 3.1 \times 10^{-2}\ ,
\end{array}
\end{equation}
where each anomalous magnetic moment is normalized to $e/2 m_q$, where 
$m_q$ is the so-called current-quark mass of each quark.
(We take $m_u=5$ MeV, $m_d=10$ MeV, $m_s=200$ MeV, $m_c=1.3$ GeV, 
$m_b=4.3$ GeV.)
The bounds on anomalous magnetic moments that we 
obtain for the other fermions are much weaker than the ones available 
in the literature (see \cite{KHayes,rem2} and references therein)
and so we do not quote them. 

For the EDM's of fermions other than the tau we may use
the restrictions from the corresponding $Z$ partial width
to constrain them. However, all the limits we obtain are much weaker 
than the ones reached by other methods  
(see \cite{KHayes,rem2} and references therein) except for
the $\nu_\tau$ EDM. Here we find
\begin{equation}
\label{dnutau}
|d(\nu_\tau)| \leq 5.2 \times 10^{-17}\ e\, \mbox{cm} \ \ \ \ (2\sigma ) \  .
\end{equation}

\section*{Acknowledgments}

We acknowledge financial support from the CICYT AEN95-0815 Research Project
and from the Theoretical Astroparticle Network under the EEC Contract 
No. CHRX-CT93-0120 (Direction Generale 12 COMA).
We thank F. Teubert and M. Mart\'{\i}nez for their comments and helpful 
collaboration in providing the most recent LEP1 data.
We are indebted to M.~J.~Lavelle for a critical reading of the manuscript.
R.~E.~acknowledges financial support by an F.~P.~I. grant from the Universitat 
Aut\`onoma de Barcelona.
E.~M.~ack\-nowledges financial support from the Direcci\'on General de 
Investigaci\'on Cient\'{\i}fica y Ense\~nanza Superior (DGICYES).
E.~M.~is grateful to Michel Fontannaz, Lluis Oliver, Olivier P\`ene, and 
to the Laboratoire de Physique Th\'eorique et Hautes Energies of the 
Universit\'e Paris XI for their hospitality while part of this work was 
performed.


\bigskip

\begin{thebibliography}{99}
\bibitem{MPerl} 
M. L. Perl, {\sl Rep. Prog. Phys.} {\bf 55} (1992) 653;\\
S. Gentile and M. Pohl, CERN-PPE-95-147 (1995),
(submitted to {\sl Phys. Rep.} {\bf C});\\ 
M. W. Gr\"unewald, HUB-IEP-95-13 (1995), (submitted to {\sl Phys. Scripta}). 
\bibitem{KHayes}
{\it Review of Particle Properties}, {\sl Phys. Rev.} {\bf D54} (1996) 1.
\bibitem{Samuel}  
M. A. Samuel, G. Li and R. Mendel, {\sl Phys. Rev. Lett.} 
{\bf 67} (1991) 668.
\bibitem{Donoghue} 
J. Donoghue, {\sl Phys. Rev.} {\bf D8} (1978) 1632.
\bibitem{rem}
R. Escribano and E. Mass\'o, {\sl Phys. Lett.} {\bf B301} (1993) 419.
\bibitem{CDF} 
F. Abe {\it et al.}, CDF Collaboration, {\sl Phys. Rev.} {\bf D43}
(1991) 2070.
\bibitem{CDF&D0ultim}
F. Abe {\it et al.}, CDF Collaboration, {\sl Phys. Rev. Lett.} {\bf 74}
(1995) 2626;\\
S. Abachi {\it et al.}, D0 Collaboration, {\sl Phys. Rev.} {\bf D52}
(1995) 4877.
\bibitem{mtCDF&D0}
P. Grannis, {\it Measurement of the mass of the Top Quark from CDF/D0},
talk presented at ICHEP96, Warsaw, 25--31 July 1996.
\bibitem{LEPEWWG}
M. W. Gr\"unewald, {\it Combined Precision Electroweak Results from LEP/SLC},
talk presented at ICHEP96, Warsaw, 25--31 July 1996.
\bibitem{rem2}
R. Escribano and E. Mass\'o, {\sl Nucl. Phys.} {\bf B429} (1994) 19.
\bibitem{listope}
C. J. C. Burges and H. J. Schnitzer, {\sl Nucl. Phys}
{\bf 228B} (1983) 464;\\
C. N. Leung, S. T. Love and S. Rao, {\sl Z. Phys.} {\bf C31} (1986) 433;\\
W. Buchm\"uller and D. Wyler, {\sl Nucl. Phys.}
{\bf 268B} (1986) 621.
\bibitem{bhm} 
BHM Electroweak library: G. Burgers, W. Hollik and M. Mart\'{\i}nez, 
Fortran program based on the Proceedings of the Workshop
on Z Physics at LEP I, CERN Report 89-08 Vol. I, 7. These computer codes
have been recently been upgraded by including the results of the
{\it Reports of the working group on precision calculations for the
Z resonance}, eds. D. Bardin, W. Hollik and G. Passarino, CERN Yellow Report
95-03.
\bibitem{limitsCP}
R. Akers {\it et al.}\,OPAL collaboration, 
{\sl Z. Phys.} {\bf C66} (1995) 31;\\
A. Stahl, {\sl Nucl. Phys. Proc. Suppl.} {\bf 40} (1995) 505.
\bibitem{millorlimitCP}
L. Silvestris, {\it A Test of CP-Invariance in $Z\rightarrow \tau^+ \tau^-$},
talk presented at ICHEP96, Warsaw, 25--31 July 1996.
\end{thebibliography}
\end{document}